\documentclass[letter]{aa} 

\usepackage{graphicx}
\usepackage{natbib}
\usepackage{xcolor}

\usepackage{txfonts}
\newcommand{\name}{HZ4}
\newcommand{\cii}{\mbox{[\ion{C}{ii}]}}
\newcommand{\hi}{\mbox{\ion{H}{i}}}

\begin{document}

   \title{Kiloparsec view of a typical star-forming galaxy when the Universe was $\sim1$~Gyr old}

   \subtitle{II. Regular rotating disk and evidence for baryon dominance on galactic scales}

  \author{R. Herrera-Camus\inst{1},
  	N. M. F{\"o}rster Schreiber\inst{2},
  	S. H. Price\inst{2},
  	H. {\"U}bler\inst{3,4},
    A. D. Bolatto\inst{5},
    R. L. Davies\inst{6,7},
    D. Fisher\inst{6,7},
    R. Genzel\inst{2},
    D. Lutz\inst{2},
    T. Naab\inst{8},
    A. Nestor\inst{9},
    T. Shimizu\inst{2},
    A. Sternberg\inst{2,9,10},
    L. Tacconi\inst{2},
    \and
    K. Tadaki\inst{11}
	 }
	\authorrunning{Herrera-Camus et al.}

   \institute{Departamento de Astronom\'ia, Universidad de Concepci\'on, Barrio 			Universitario, Concepci\'on, Chile\\
              \email{rhc@astro-udec.cl}
              \label{1}
              \and
          Max-Planck-Institut f\"ur extraterrestische Physik (MPE), Giessenbachstr., D-85748 Garching, Germany
	\label{2}
	\and
		Cavendish Laboratory, University of Cambridge, 19 J.J. Thomson Avenue, Cambridge CB3 0HE, UK
  		\label{3}
  	\and
		Kavli Institute for Cosmology, University of Cambridge, Madingley Road, Cambridge CB3 0HA, UK
  		\label{4}
	\and
	Department of Astronomy, University of Maryland, 
	College Park, MD 20742, USA 
	\label{6}
	\and
	Centre for Astrophysics and Supercomputing, Swinburne Univ. of Technology, P.O. Box 218, Hawthorn, VIC 3122, Australia
	\label{7}
	\and
	ARC Centre of Excellence for All Sky Astrophysics in 3 Dimensions (ASTRO 3D), Australia
	\label{5} 
	\and
	Max-Planck Institute for Astrophysics, Karl Schwarzschildstrasse 1, D-85748 Garching, Germany
	\label{8}
	\and
	School of Physics and Astronomy, Tel Aviv University, Ramat Aviv 69978, Israel
	\label{9}
	\and
    Center for Computational Astrophysics, 162 5th Ave., New York, NY, 10010
	\label{10}	
	\and
	National Astronomical Observatory of Japan, 2-21-1 Osawa, Mitaka, Tokyo 181-8588, Japan
	\label{11}
             }

   \date{Received ; accepted }

 
  \abstract
  {
  We present a kinematic analysis of the main-sequence galaxy \name\ at $z=5.5$. Our study is based on deep, spatially resolved observations of the \cii~158~$\mu$m transition obtained with the Atacama Large Millimeter/Submillimeter Array (ALMA). From the combined analysis of the disk morphology, the two-dimensional velocity structure, and forward-modeling of the one-dimensional velocity and velocity dispersion profiles, we conclude that \name\ has a regular rotating disk in place. The intrinsic velocity dispersion in \name\ is high ($\sigma_{0}=65.8^{+2.9}_{-3.3}$~km~s$^{-1}$), and the ratio between the rotational velocity and the intrinsic velocity dispersion is $V_{\rm rot}/\sigma_{0}=2.2$. These values are consistent with the expectations from the trends of increasing $\sigma_{0}$ and decreasing $V_{\rm rot}/\sigma_{0}$ as a function of redshift observed in main-sequence galaxies up to $z\approx4$. Galaxy evolution models suggest that the high level of turbulence observed in \name\ can only be achieved if, in addition to stellar feedback, there is radial transport of gas within the disk. Finally, we find that \name\ is baryon dominated on galactic scales ($\lesssim2\times R_{\rm e}$), with a dark matter fraction at one effective radius of $f_{\rm DM}(R_{\rm e})=0.41^{+0.25}_{-0.22}$. This value is comparable to the dark matter fractions found in lower redshift galaxies that could be the descendants of \name: massive ($M_{\star}\approx10^{11}~M_{\odot}$), star-forming galaxies at $z\sim2$, and passive, early type galaxies at $z\approx0$.
  }


   \maketitle
%

\section{Introduction}

The study of galaxy kinematics can provide answers to some of the most fundamental questions about galaxy formation and evolution: How and when do galaxies form their disks? What is the distribution of baryons and dark matter within galaxies? What is the dynamical state of galaxy disks? From early kinematic analyses of nearby galaxies \cite[e.g.,][]{rhc_rubin70,rhc_sofue01}, to more recent studies of high redshift systems thanks to powerful interferometers and integral-field units (IFU) systems, especially with adaptive optics \citep[e.g.,][]{rhc_glazebrook13,rhc_nfs20}, we can now start connecting the kinematic properties of the very early galaxies with their likely descendants.

Based on deep, spatially resolved observations of the H$\alpha$ and CO transitions in star-forming galaxies between $0\lesssim z \lesssim3$, we have learned that: (1) the amount of turbulence in the interstellar medium (ISM) ---measured by the gas intrinsic velocity dispersion---  increases as a function of redshift \citep[e.g.,][]{rhc_genzel06,rhc_nfs06,rhc_cresci09,rhc_kassin12,rhc_stott16,rhc_johnson18, rhc_uebler19}, (2) galaxy disks are more dynamically turbulent (or {\it dynamically hot}) at higher redshift, quantified by the ratio between the rotational velocity and the intrinsic velocity dispersion \citep[e.g.,][]{rhc_law09,rhc_nfs09,rhc_simons17,rhc_wisnioski15,rhc_wisnioski19}, and (3) more than half of the massive, star-forming galaxies at $z\sim1-2$ are baryon dominated on galactic scales \citep[e.g.,][]{rhc_wuyts16,rhc_lang17,rhc_uebler18,rhc_genzel17,rhc_genzel20,rhc_price21}. These observational results point to the importance of stellar feedback, clump formation, and gas transport in the evolution of galaxies \citep[e.g.,][]{rhc_dekel09,rhc_dekel14,rhc_ostriker11,rhc_bournaud14,rhc_krumholz18}.

The kinematic properties of star-forming galaxies at $z\gtrsim4$ remain relatively unexplored. This situation, however, is rapidly changing, thanks to observations of the \cii~158~$\mu$m fine-structure line. The \cii\ transition originates from the collisional excitation of C$^+$ ions by electrons, H$_2$ molecules, and ---to the advantage of kinematic studies--- H atoms. For now, observations of the \hi~21~cm line in high redshift galaxies are out of reach. This makes the \cii\ transition an excellent alternative to trace the outer disk of distant galaxies \citep[e.g.,][]{rhc_madden93,rhc_deblok16,rhc_fujimoto19,rhc_fujimoto20,rhc_rhc21}.

Between $z\sim4-6$, a morpho-kinematic analysis of 29 star-forming galaxies based on \cii\ line observations with modest angular resolution ($\sim1\arcsec$ or $\sim6.5$~kpc at $z=5$), indicates a diversity of kinematic types, including rotators, dispersion dominated systems, and mergers \citep{rhc_jones21}. This result is confirmed by the handful of spatially-resolved \cii\ observations available of typical star-forming galaxies probing the main-sequence population in the $M_{\star}-{\it SFR}$ plane at $z\gtrsim4$. Evidence for rotating disks has been found in two Lyman break galaxies at $z\approx6.8$ \citep{rhc_smit18}, one absorption-selected galaxy at $z=4.2$ \citep{rhc_neeleman20}, and a handful of dusty, star-forming galaxies at $z\approx4$ \citep{rhc_rizzo20,rhc_rizzo21}. There are also examples of complex kinematics including a potential interaction between three systems \citep[LBG-1; ][]{rhc_riechers13}, and two massive, star-forming galaxies in the process of merging \citep{rhc_ginolfi20b}. In the near future, ALMA, NOEMA and the {\it James Webb Space Telescope (JWST)} will continue contributing to the rapid increase of kinematic studies of high-redshift, star-forming galaxies.

In this letter, we present one of the most complete kinematic analyses to date of a main-sequence, star-forming galaxy at $z\approx5$. For this, we use  deep, spatially-resolved \cii\ line observations of the star-forming galaxy \name\ at $z=5.5$ \cite[$M_{\star}=10^{10.15}~M_{\odot}$, ${\rm SFR}=40.7~M_{\odot}~yr^{-1}$; ][]{rhc_faisst20}. The observations and the analysis of the morphology, ISM, extended emission (or ``halo''), and outflow properties of \name\ are presented in \cite{rhc_rhc21} (hereafter Paper~I).


\section{Observations and data reduction} \label{sec:data}

For a detailed description of the observations and data reduction we refer to Paper~I. \name\ was observed with ALMA for a total of 8.4~hr (4.7~hr on source) as part of project 2018.1.01605.S (PI Herrera-Camus). The observations were carried out in the C34-4 configuration in Band~7. The data were processed using the Common Astronomy Software Applications package \citep[CASA;][]{rhc_casa} version 5.6.2. Using the CASA task \texttt{tclean}, Briggs weighting (robust=+0.5), and the Multi-Scale Clean algorithm \citep{rhc_cornwell08}, we generated a \cii\ cube with a synthesized beam size of $\theta_{\rm beam}=0.39\arcsec\times0.34\arcsec$ at a position angle of ${\rm PA}=-72.4^{\rm o}$. The velocity resolution of the \cii\ cube is 16~km~s$^{-1}$. This is at least a factor $\sim4$ higher than the best spectral resolution that can be achieved in near-infrared observations, so the line spread effect is negligible. The rms noise for the \cii\ cube was 0.15~mJy~beam$^{-1}$ in 16~km~s$^{-1}$ channels.

\begin{figure*}
\centering
\includegraphics{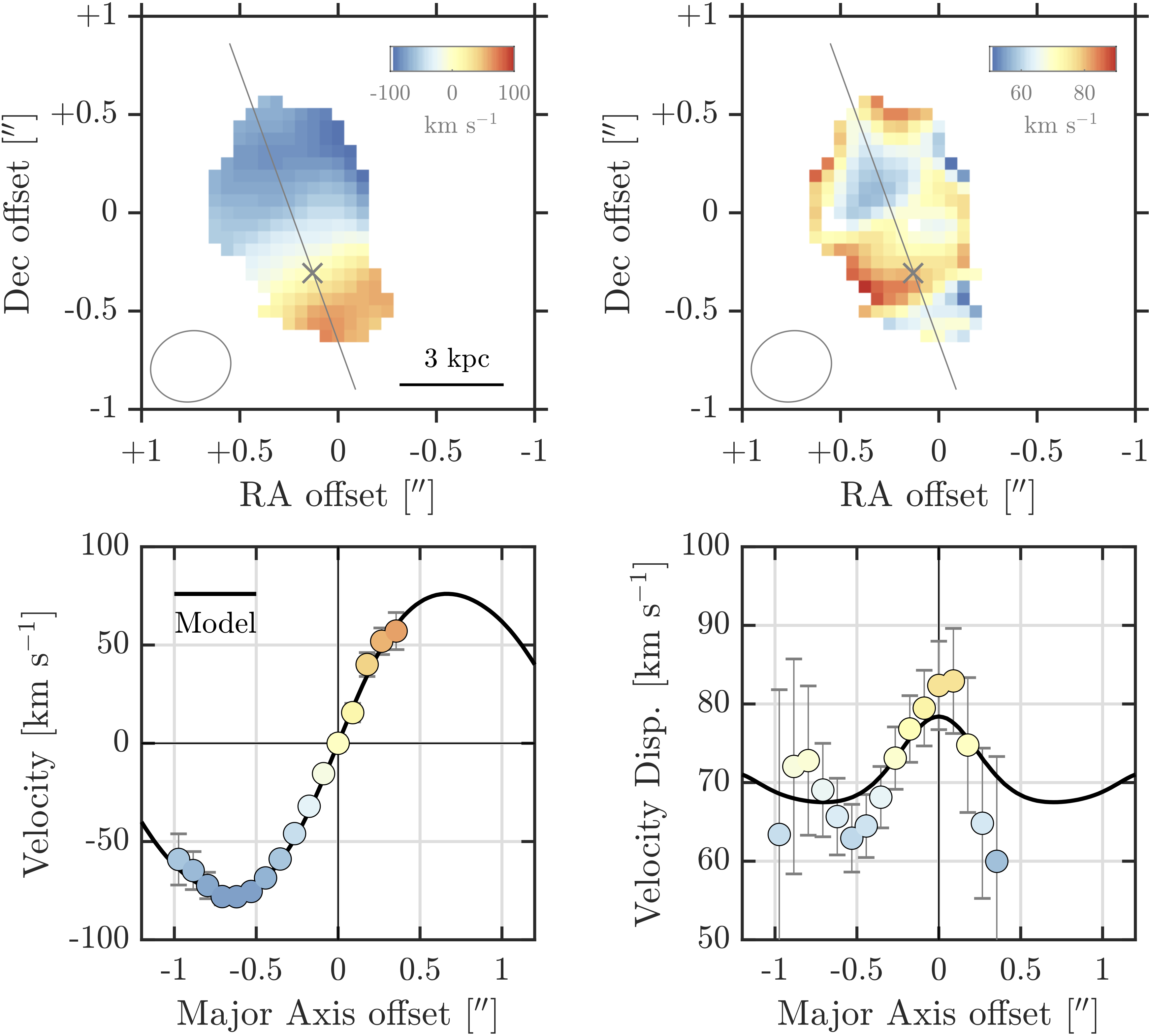}
\caption{{\it (Top)}  \cii\ velocity field (left) and velocity dispersion map (right) of \name. The ALMA synthesized beam ($\theta=0.39\arcsec \times 0.34\arcsec$) is shown in the bottom-left corner. {\it (Bottom)} Rotation curve and velocity dispersion profiles extracted employing a pseudo-slit oriented along the kinematic major axis shown by the gray solid line in the upper panels, where the kinematic center is shown with a gray cross. The solid black line in both panels represent the best-fit DYSMAL model beam-convolved to the observed space. \label{fig:mom+RC}
}
\end{figure*}
 
\section{Results and analysis} \label{sec:results}

\subsection{Basic kinematic properties}\label{sec:kinprop}

We created 2D kinematic maps of \name\ by fitting a Gaussian profile to the \cii\ line emission in each pixel and accounting for the systemic velocity. Fig.~\ref{fig:mom+RC} shows the resulting velocity field (first moment; top-left) and velocity dispersion map (second moment; top right). The kinematic center is defined from the combination of the centrally peaked velocity dispersion and the location of steepest
gradient in the velocity field, resulting in the position R.A. +09:58:28.5 and Dec. +02:03:06.3 (shown with a gray cross in Fig.~\ref{fig:mom+RC}). The major kinematic axis is determined from the 2D velocity field as the angle at which the radial velocity profile centered in the kinematic center includes the largest observed velocity difference. This results in a position angle (PA) of $200^{\circ}$ measured anti-clockwise from north to the receding side of the galaxy (or $20^{\circ}$ measured east of north; shown with a gray line in Fig.~\ref{fig:mom+RC}).



Based on a an initial kinematic analysis, we find that \name\ fulfills at minimum four main characteristics of a rotating disk as described in \cite{rhc_nfs20}. First, there is a smooth velocity gradient observed across the galaxy. 
Second, the kinematic major axis is aligned with the \cii\ morphological major axis (which has a ${\rm PA}=17.7^{\circ}$ measured east from north, Paper~I). Third, \name\ has a centrally peaked velocity dispersion distribution.
Fourth, the kinematic center and the morphological center (see Paper I) are spatially coincident, separated by a projected distance of only $\sim0.2\arcsec$ ($\sim1$~kpc), which is smaller than the beam size. These basic kinematic properties suggest that \name\ has a smooth, rotating disk in place at $z\approx5.5$. The modeling and discussion of the kinematics properties of \name\ are presented in Section~\ref{sec:analysis}.


\subsection{Rotation curve and velocity dispersion profile}

\begin{figure*}
\centering
\includegraphics[scale=0.23]{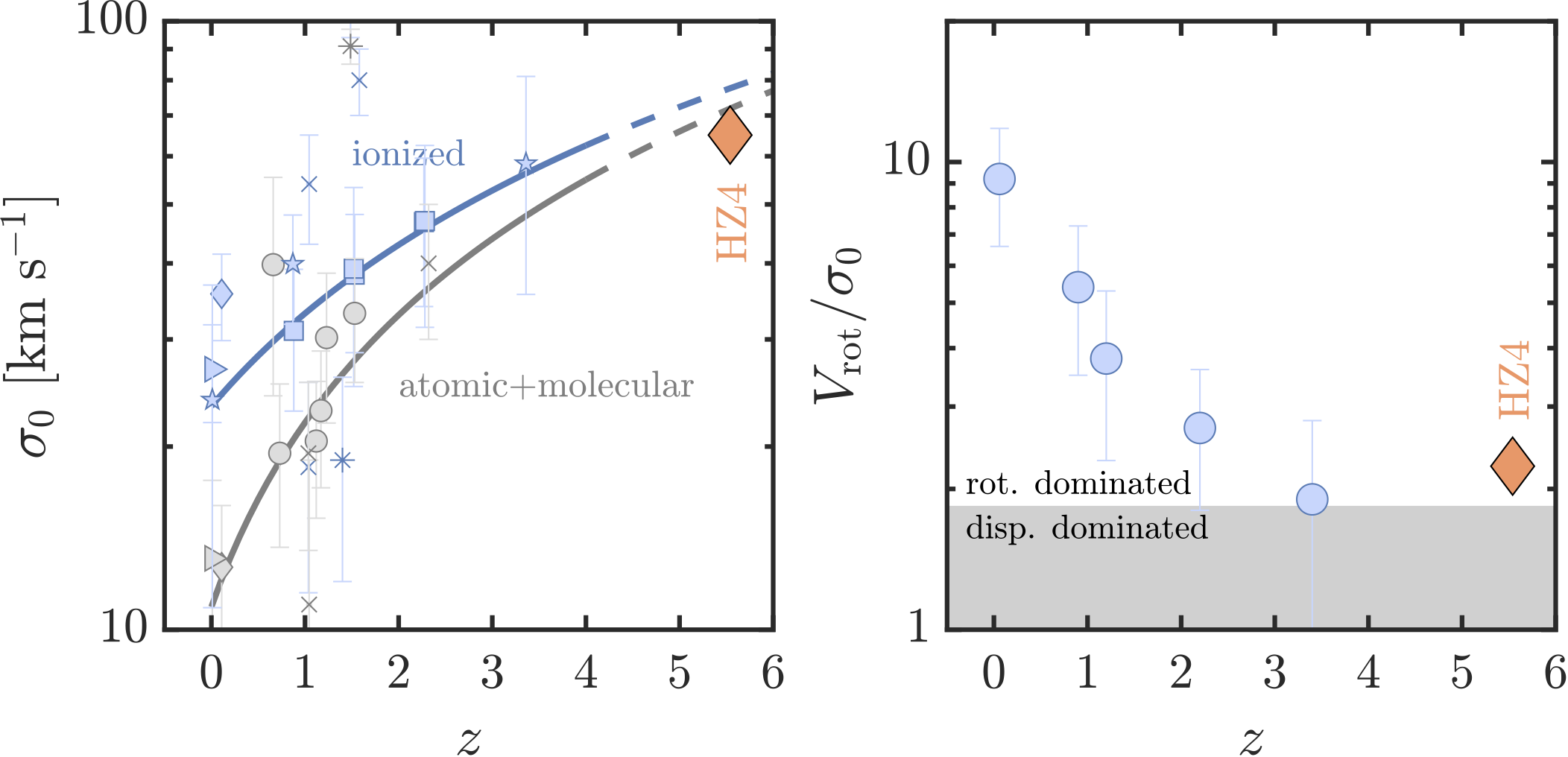}
\caption{Evolution of the intrinsic velocity dispersion ($\sigma_0$; left) and the disk dynamical support ($V_{\rm rot}/\sigma_0$; right) as a function of redshift. \name\ at $z\approx5$ is shown as an orange diamond. In the left panel, the circles show average values from different surveys of ionized (blue) and atomic and/or molecular gas (gray). These include: DYNAMO \citep[diamonds; ][]{rhc_fisher19,rhc_girard21}, HERACLES, THINGS and EDGE \citep[triangles; ][]{rhc_leroy08,rhc_leroy09,rhc_mogotsi16,rhc_bolatto17,rhc_levy18}, KMOS$^{\rm 3D}$ and SINS/zC-SINF \citep[squares; ][]{rhc_nfs06,rhc_nfs09,rhc_wisnioski15,rhc_wisnioski19,rhc_uebler19}, PHIBSS \citep[circles; ][]{rhc_tacconi13,rhc_freundlich19}, and GHASP/KDS/KROSS \citep[stars; ][]{rhc_epinat10,rhc_stott16,rhc_turner17,rhc_johnson18}. We also add individual measurements from lensed systems (crosses) from \cite{rhc_swinbank11} and \cite{rhc_girard19}, and unlensed systems (asterisks) from \cite{rhc_molina19} and \cite{rhc_uebler18}. The solid lines show the best-fit relations to the observations compiled by \cite{rhc_uebler19} up to $z\approx3.5$. In the right panel, the circles show the average values for $V_{\rm rot}/\sigma_0$ from the ionized gas measured and compiled by \cite{rhc_wisnioski15}. The gray box shows the region below $V_{\rm rot}/\sigma_0=\sqrt{3.36}$ where the contribution to the dynamical support of the disk by random motions starts to dominate \citep[e.g., ][]{rhc_nfs20}.
\label{fig:evol}
}
\end{figure*}

We constructed the rotation curve and velocity dispersion profile from the \cii\ cube by placing 0.4\arcsec\ diameter apertures ($\sim$beam size) along the kinematic major axis, separated every $\sim$0.1\arcsec. We emphasize that for the purpose of our analysis, the major-axis information provides the strongest constraints on the disk models we use (see Section \ref{sec:dysmal}). The resulting velocity and dispersion profiles are shown in the bottom panels of Fig.~\ref{fig:mom+RC}.

The rotation curve of \name\ is more extended in the north-east (approaching) side, and reaches out to $\sim1\arcsec$. This corresponds to a projected physical distance of $\sim$6~kpc, or $\sim2\times$ the effective radius $R_{\rm e}$ (as measured from fitting an exponential disk profile to the integrated \cii\ emission, Paper I). The rotation curve peaks at an observed projected velocity of 80~km~s$^{-1}$ at a distance of $\sim1.2\times R_{\rm e}$ from the center, and then drops down to 60~km~s$^{-1}$ at $\sim2\times R_{\rm e}$. Dropping rotation curves are also observed in massive, star-forming galaxies at $z\sim1-2$ \citep[][]{rhc_genzel17,rhc_lang17,rhc_genzel20}. As we discuss in more detail in Section \ref{sec:analysis}, these can be interpreted as the combination of low central dark matter fractions and/or the effect of asymmetric drift due to high intrinsic velocity dispersions \cite[e.g.,][]{rhc_burkert10}.

\subsection{Kinematic modeling}\label{sec:dysmal}

We simultaneously model the velocity and velocity dispersion profiles extracted along the major kinematic axis using an updated version of the fully 3D parametric code DYSMAL \citep{rhc_cresci09,rhc_davies11,rhc_wuyts16,rhc_uebler18,rhc_price21}. DYSMAL uses a forward modeling approach and  accounts for all important observational effects, including spatial beam smearing. The code creates a 3D mass model, forward model to produce a 3D spectroscopy cube, from which we extract the 1D profiles in the same fashion as the observations. These are then compared to the data using an MCMC sampling procedure \citep[{\tt EMCEE}, ][]{rhc_foreman-mackey13}. For a detailed and updated description of the DYSMAL code, we refer to \cite{rhc_price21}.

We model \name\ as a thick, turbulent disk embedded in a dark matter halo that follows a NFW profile \citep{rhc_navarro96} and has a halo mass $M_{\rm halo}$ and a concentration parameter $c_{\rm halo}$. For the galaxy, we assume an exponential disk with effective radius $R_{\rm e}$. These are reasonable assumptions based on the analysis of the integrated \cii\ line emission in Paper~I. DYSMAL calculations are based on the assumed shape of the baryon distribution. Then the code differentiate between the circular velocity from the baryon distribution and the NFW profile. We also assume an intrinsic velocity dispersion ($\sigma_{0}$) that is constant and isotropic throughout the disk. The model includes an asymmetric drift correction to the model circular velocity due to pressure support following \cite{rhc_burkert10}. 

Free parameters in our modeling are the total baryonic mass ($M_{\rm bar}$), the effective radius $R_{\rm e}$, the intrinsic velocity dispersion $\sigma_{0}$, and the enclosed dark matter fraction within one effective radius ($f_{\rm DM}(R_{\rm e})$). The latter is measured as the squared ratio between the dark matter ($v_{\rm circ,DM}$) and total ($v_{\rm circ,tot}$) intrinsic circular velocities measured at $R_{\rm e}$, i.e., $f_{\rm DM}(R_{\rm e})=v_{\rm circ,DM}^2( R_{\rm e})/v_{\rm circ,tot}^2(R_{\rm e})$. For the total baryonic mass, we choose a Gaussian prior bounded in the range $\log_{10}(M_{\rm bar}/M_{\odot}) \in [10,11.5]$~dex with a standard deviation of 0.2~dex, and centered at the expected baryonic mass derived from the sum of the stellar \citep[$M_{\star}=10^{10.15}~M_{\odot}$;][]{rhc_faisst20} and gas mass assuming a gas fraction of $M_{\rm gas}/M_{\star}=0.75$ \citep[e.g.,][]{rhc_dessauges-zavadsky20}. For the effective radius we also assume a Gaussian prior bounded in the range $ R_{\rm e}\in[2.5,4.5]$~kpc and centered at a value of $R_{\rm e}=3.4$~kpc following the results from Paper I. For the intrinsic velocity dispersion and the dark matter fraction, we adopt flat bounded priors of $\sigma_{0}\in[20,100]$~km~s$^{-1}$ and $f_{\rm DM}(R_{\rm e})\in[0,1]$, respectively.

Given the difficulty to simultaneously fit for more parameters due to the resolution and S/N of the data, we fix: (1) the kinematic center and position angle, as described in Section~\ref{sec:kinprop}; (2) the inclination to $i=52^{\circ}$, based on the morphology of the ALMA \cii\ line data (Paper~I) and assuming a ratio of scale height to scale length of thickness $=0.15$; and (3) the halo concentration parameter, $c_{\rm halo}=2.3$, typical for the redshift and halo mass of \name\ according to \cite{rhc_dutton14}.

The MCMC code is run using 800 walkers, a burn-in phase of 800 steps, and a running phase of 5000 steps. Fig.~\ref{fig:corner} in Appendix~\ref{sec:appendix} shows the MCMC sampling of the joint posterior probability distribution (or ``corner plot''). 
The maximum a posteriori values of the free parameters (found by jointly analyzing the posteriors of all free parameters) are: 
$\log_{10}(M_{\rm bar}/M_{\odot})=10.38^{+0.25}_{-0.13}$, $R_{\rm e}=3.4^{+0.5}_{-0.3}$~kpc, $f_{\rm DM}(R_{\rm e})=0.41^{+0.25}_{-0.22}$, and $\sigma_{0}=65.8^{+2.9}_{-3.3}$~km~s$^{-1}$. We consider these parameters as the model that best describe our data. The corresponding model velocity and velocity dispersion profiles in the observed space are shown as black solid lines in Fig.~\ref{fig:mom+RC}. In addition, Fig.~\ref{fig:pv} in Appendix~\ref{sec:pv-appendix} shows the position-velocity diagram of \name\ and the best-fit DYSMAL model cube extracted along the major and minor kinematic axes.

The model-derived effective radius is comparable to that measured from the exponential profile fit to the integrated \cii\ emission (Paper~I), and the resulting baryonic mass translates into a high model-based gas fraction of $M_{\rm gas}/M_{\star}=(M_{\rm bar}-M_{\star})/M_{\star}=0.7$,
comparable to the mean \cii-based gas mass fraction measured in main-sequence, star-forming galaxies between $5.1<z<5.9$ \citep{rhc_dessauges-zavadsky20}.

If we vary the assumption of the inclination by $\pm10\%$, we observe a similar variation in the derived baryonic mass and the dark-matter fraction. Also, if we fix $R_{\rm e}=3.4$~kpc (Paper~I), and let the inclination free with a Gaussian prior centered at $i=52^{\circ}$ and bounded in the range $i\in [30^{\circ},70^{\circ}]$, then the model-derived inclination results $i=51.9^{+5.6}_{-10.7}$ deg, an the dark-matter fraction is $f_{\rm DM}(R_{\rm e})=0.45^{+0.19}_{-0.28}$, consistent with the results from the fixed-inclination model.


\section{Discussion} \label{sec:analysis}

\begin{figure*}
\centering
\includegraphics[scale=0.23]{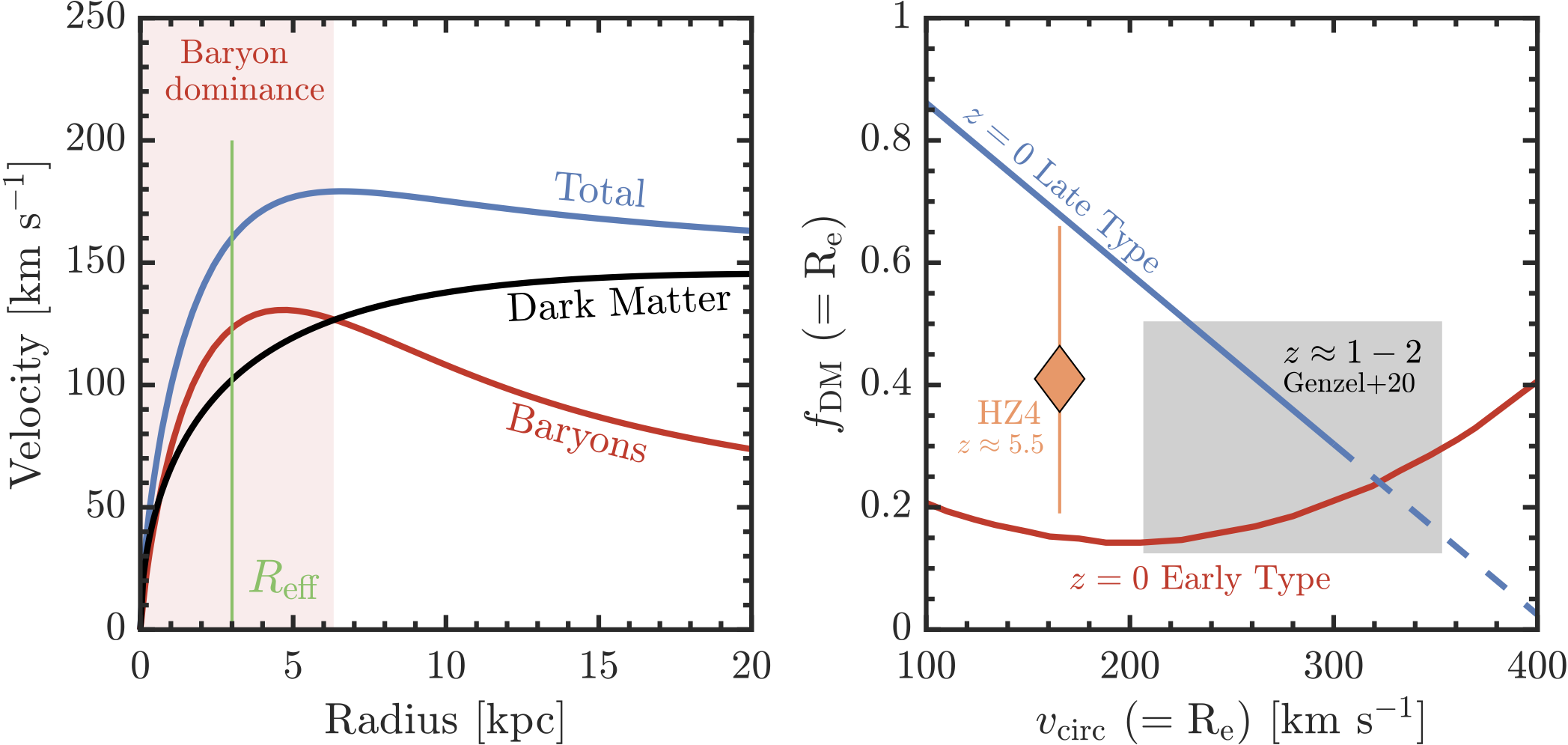}
\caption{({\it Left)} Intrinsic rotation curve (corrected by inclination) of \name\ for baryons (red), dark matter (black), and total (blue) from the best-fit model. The green vertical line shows the $ R_{\rm e}$, and the red colored box represents the spatial scales of baryon dominance in \name. {(\it Right)} Dark matter fraction within $ R_{\rm e}$ of \name\ (orange diamond), main-sequence, star-forming galaxies at $z\approx1-2$ \citep[gray box; ][]{rhc_genzel20}, and the best linear fit to $z\sim0$ late type galaxies (blue line), and $z\sim0$ ATLAS-3D early type galaxies \citep[red curve; ][]{rhc_cappellari16}.
\label{fig:model}
}
\end{figure*}

\subsection{Velocity dispersion and disk stability over cosmic time}

The intrinsic gas velocity dispersion of main-sequence star-forming galaxies has been observed to increase as a function of redshift at least up to $z\sim4$ \citep[e.g,][]{rhc_nfs06,rhc_kassin07,rhc_genzel11,rhc_wisnioski15,rhc_turner17,rhc_simons17, rhc_johnson18,rhc_uebler19,rhc_livermore15}. This evolution is consistent with the observed increase in the molecular gas fractions with redshift \citep[e.g.,][]{rhc_tacconi20}: if galaxy disks grow in a marginally stable equilibrium state with Toomre parameter $Q\sim1$, then it is expected that $V_{\rm rot}/\sigma_0\propto 1/f_{\rm gas}$ \citep[e.g.,][]{rhc_genzel11,rhc_glazebrook13,rhc_wisnioski15}.

Fig.~\ref{fig:evol} shows the evolution of $\sigma_{0}$ observed in main-sequence galaxies as a function of redshift. The points correspond to the averages of several galaxy surveys and individual measurements in the ionized and atomic and/or molecular gas, measured and/or compiled by \cite{rhc_uebler19}. The high intrinsic velocity dispersion inferred from the kinematic modeling of \name\ ($\sigma_{0}\approx65$~km~s$^{-1}$) is consistent with the extrapolation up to redshift $z\approx5$ of the scaling relations derived for main-sequence galaxies. This is in line with the expectations from the observed increase of the cold gas fraction with redshift \citep[][]{rhc_tacconi18,rhc_tacconi20}, of which \name\ is no exception with a cold gas mass fraction of $M_{\rm gas}/M_{\star}=0.7$ at $z\approx5.5$.

$V_{\rm rot}/\sigma_{0}$ is a measure of the dynamical support contributed by rotation versus random motions. We infer from the kinematic modeling a value of $V_{\rm rot}/\sigma_{0}=2.2$. As the right panel of Fig.~\ref{fig:evol} shows, this value is consistent with the evolution observed in main-sequence galaxies \citep[][and references therein]{rhc_wisnioski15,rhc_wisnioski19}, and is above the threshold assumed for equal contribution to the dynamical support of disks from rotation and random motions \citep[$V_{\rm rot}/\sigma_{0}\approx \sqrt{3.36}$; e.g.,][]{rhc_nfs20}.  The $V_{\rm rot}/\sigma_{0}$ value measured in \name, combined with the basic kinematic properties discussed in Section~\ref{sec:kinprop}, provide one of the most compelling pieces of evidence to date of the existence of regular rotating disks when the Universe was only $\sim1$~Gyr old. 

To date, there are a handful of studies that find evidence for rotating disks in the early universe \citep[e.g.,][]{rhc_smit18,rhc_neeleman20,rhc_lelli21,rhc_rizzo20,rhc_rizzo21}. Interestingly, studies from \cite{rhc_lelli21}, \cite{rhc_rizzo21} and \cite{rhc_fraternali21} conclude that the dusty starbursts in their samples are dynamically cold, with $V_{\rm rot}/\sigma_{0}$ values in the $\sim7-30$ range. As opposed to the case of \name, these values are significantly above the expectations from the observed evolution of $V_{\rm rot}/\sigma_{0}$ for main-sequence galaxies in the range $z\approx0-4$ (see Fig.~\ref{fig:evol}). These differences could indeed reflect in part the different nature of the sources, with most of the \cite{rhc_rizzo21}, \cite{rhc_lelli21} and \cite{rhc_fraternali21} targets (6 of 8) lying at least an order of magnitude above the main-sequence and being typically very compact compared to \name\ and other galaxies close to the main-sequence.  We caution however that direct comparisons are also complicated by the different modeling approaches between the studies, and notably in the $\sigma_{0}$ parameterization (radially constant in this work, exponentially decreasing in the \cite{rhc_rizzo20,rhc_rizzo21} work).

\subsection{Dark matter fraction on galactic scales}

Recent observational evidence shows that a large fraction of massive, star-forming galaxies at $z\sim1-2$ are strongly baryon-dominated on galactic scales \citep[e.g.,][]{rhc_wuyts16,rhc_uebler18,rhc_genzel17,rhc_genzel20,rhc_price21}. Measuring dark matter fractions requires well-sampled rotation curves that extend beyond the effective and turnover radius. Up to $z\sim2$, these outer rotation curves have been obtained mainly by observations of the H$\alpha$ and CO transitions. In the case of \name, our deep \cii\ line observations allow us to trace the rotation curve beyond the turnover point ($\sim1.2\times R_{\rm e}$), and as far as $\sim2\times R_{\rm e}$. This enables a robust decomposition of the baryonic and dark-matter components of the rotation curve based on our DYSMAL analysis. 

Fig.~\ref{fig:model} (left) shows the intrinsic  (inclination corrected) baryonic, dark matter, and total circular velocity profile of \name. The effective radius is shown with a green vertical line. We find that \name\ is baryon-dominated within $ R_{\rm e}$, and the baryon dominance prevails out to $\sim2\times R_{\rm e}$. The right panel of Fig.~\ref{fig:model} shows the $f_{\rm DM}( R_{\rm e}$) of \name\ in context with other galaxy populations. The blue line shows the best fit to the $z\approx0$ late type galaxies\footnote{The fit to $z\approx0$ late-type galaxies from \cite{rhc_genzel20} (and references therein) corresponds to: $f_{\rm DM,late-type}=1-0.279\times(v_{\rm circ}-50.3~{\rm km ~s^{-1}})/100~{\rm km~s^{-1}}$.}. While low-mass galaxies tend to be dark-matter dominated, more massive disks have lower dark matter fractions. The red line shows the best fit to passive, early type galaxies at $z\approx0$, which are strongly baryon dominated within one $ R_{\rm e}$ \citep{rhc_cappellari16}. The gray box shows the mean ($\pm$ the standard deviation around the mean) of the dark matter fraction measured in 41 massive ($M_{*}\approx10^{10.5}-10^{11}~M_{\odot}$), main-sequence galaxies at $z\approx1-2$ \citep{rhc_genzel20}. 

At the same $v_{\rm circ}$, and noting the time difference in evolution of $\sim12.5$~Gyr, \name\ has a dark matter fraction in between the values from late-type (dark matter dominated) and early-type (baryon dominated) galaxies at $z\approx0$, and comparable to the mean value found in more massive ($M_{\star}\sim10^{11}~M_{\odot}$), star-forming galaxies at $z\approx2$. This is interesting because, from an abundance-matching perspective \citep[e.g.,][]{rhc_hill17}, the progenitors of the baryon-dominated, massive galaxies at $z\approx2$ are $M_{\star}\approx10^{10}~M_{\odot}$ galaxies at $z\approx5$, the stellar mass of \name. Could it be then that the baryon dominance observed in massive, star-forming galaxies at cosmic noon is already in place at $z\approx5$? This is a possibility, as 
the central dark-matter fraction 
of massive star-forming galaxies at $z\approx2$, and their likely early-type descendants at $z\approx0$, is comparable \citep[e.g.,][]{rhc_uebler18,rhc_genzel17,rhc_genzel20}. This result is also supported by the Illustris-TNG simulations, that find only a mild evolution with redshift of the dark-mater fraction on galactic scales for a fixed stellar mass \citep{rhc_lovell18}. In fact, TNG-galaxies at $z\approx4$ with a similar stellar mass than \name\ have a mean dark-matter fraction within one stellar half mass radius of $f_{\rm DM}\approx0.4$, and this value only increases by $10-20\%$ as galaxies grow to a stellar mass of $M_{\star}\approx10^{11}~M_{\odot}$ at $z\approx2$.

\subsection{Comparison to galaxy evolution models including feedback and gravity} 

The high intrinsic velocity dispersion in \name\ is consistent with the observed trends of increasing velocity dispersion with redshift. The turbulence in these high-$z$ galactic disks is expected to be primarily driven by thermal and momentum feedback from massive stars and supernovae \citep[e.g.,][]{rhc_ostriker11,rhc_shetty12}, clump formation, and radial gas transport \citep[e.g.,][]{rhc_dekel09,rhc_dekel14,rhc_bournaud14,rhc_krumholz18}. 

Here we investigate what processes are responsible for the high intrinsic velocity dispersion observed in \name\ using the state-of-the-art analytical model of galaxy evolution by \cite{rhc_krumholz18}. One of the advantages of the \cite{rhc_krumholz18} model is that it unifies the effect of star formation feedback and radial transport in driving the turbulence in galactic disks. We consider four model scenarios: (1) gas transport plus stellar feedback, (2) gas transport without stellar feedback, and two additional models of stellar feedback without gas transport, assuming (3) a fixed star formation efficiency per free-fall time ($\epsilon_{\rm ff}=0.015$), and (4) a fixed Toomre parameter $Q$. \footnote{As described in \cite{rhc_krumholz18}, model (3) is similar to the model described by \cite{rhc_ostriker11}, and model (4) is similar to the one proposed by \cite{rhc_f-g13}} Finally, for comparison with \name, we choose a set of parameters in the \cite{rhc_krumholz18} model that are representative of high-$z$ galaxies. These include a total circular velocity of 200~km~s$^{-1}$ measured at a distance of 5~kpc, and a molecular gas fraction of $M_{\rm gas}/M_{\star}=0.75$, in good agreement with \name.

\begin{figure}
\centering
\includegraphics[scale=0.14]{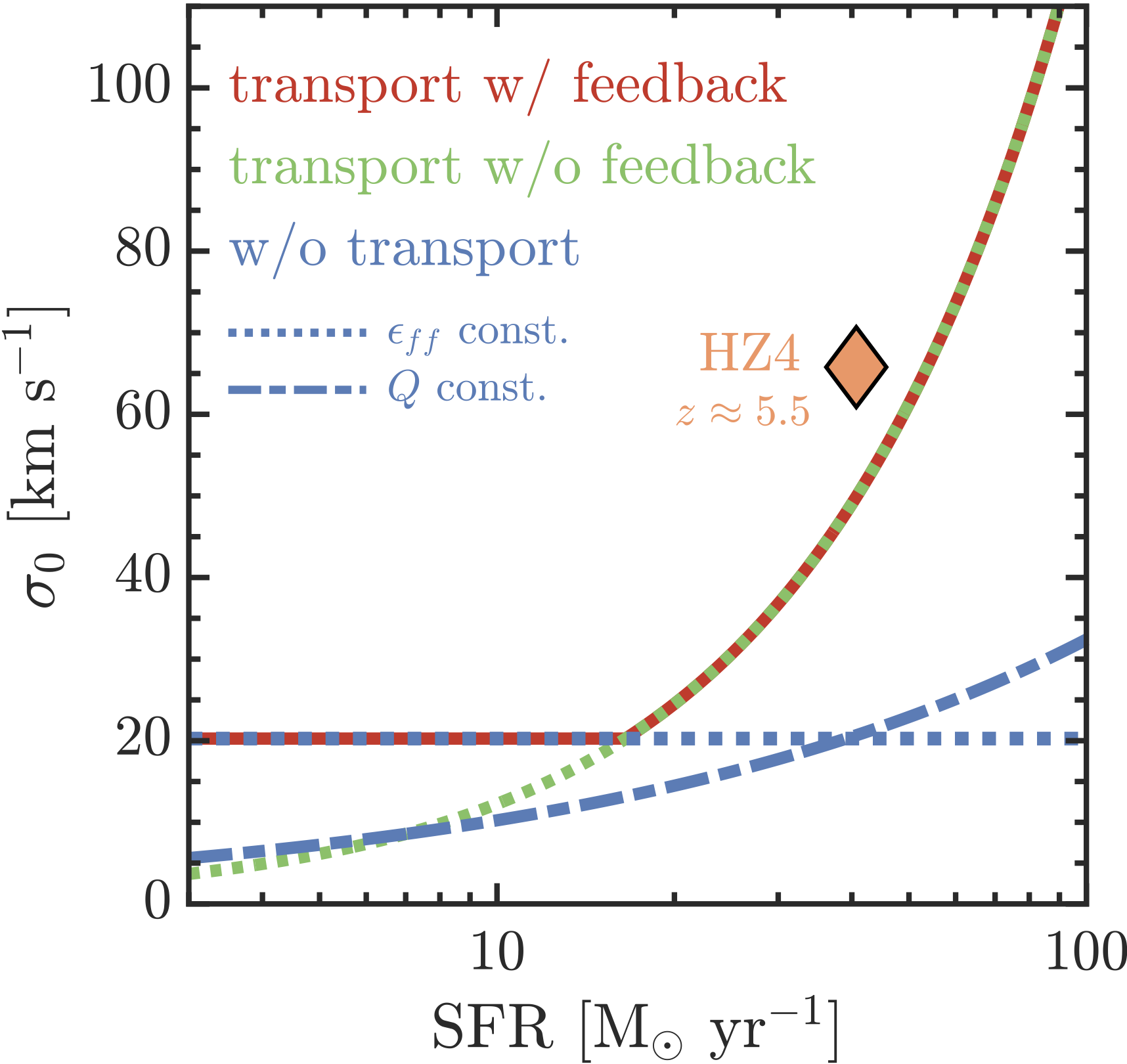}
\caption{Intrinsic velocity dispersion $\sigma_0$ as a function of star formation rate ${\rm SFR}$ for \name\ (orange diamond), and the unified models by \cite{rhc_krumholz18} for high-$z$ galaxies that consider transport+feedback (red), only transport (green), and feedback without transport (blue): the dotted and dashed lines correspond to the models with constant $\epsilon_{\rm ff}$ and constant $Q$, respectively. \label{fig:transport}
}
\end{figure}

Fig.~\ref{fig:transport} shows the comparison between the measured $\sigma_{0}$ and SFR in \name, and the predictions from the four \cite{rhc_krumholz18} models. The blue lines correspond to the models that only consider stellar feedback. In the case $\epsilon_{\rm ff}$ is fixed (dotted line), the model predicts a constant $\sigma_{0}$ of $\approx20$~km~s$^{-1}$. If $Q$ is fixed (dashed line), there is no $\sigma_{0}$ floor, but $\sigma_{0}$ only mildly increases with star formation activity. In both cases, stellar feedback alone is not capable of reproducing the high turbulence observed in \name. The situation is completely different with models that include radial transport of gas. The red and green lines show such models with and without including stellar feedback, respectively. For ${\rm SFR}\lesssim15~M_{\odot}~{\rm yr}^{-1}$, the main difference between the two models is the $\sigma_{0}$ floor of about $\approx20$~km~s$^{-1}$ in the case where stellar feedback is included. For ${\rm SFR}\gtrsim15~M_{\odot}~{\rm yr}^{-1}$, both models are identical, and $\sigma_{0}$ increases rapidly with SFR. At the SFR of \name, the radial transport models reproduce remarkably well the high $\sigma_{0}$ measured in the system.

One of the consequences of gas richness and efficient radial transport of gas and clumps inward is the formation of massive bulges, and the increase in nuclear gas outflows powered by central starbursts and active galactic nuclei \citep[e.g.,][]{rhc_elmegreen08,rhc_dekel13,rhc_bournaud14,rhc_ceverino15}. In that sense, the presence of migrating gas in \name\ suggested by the \cite{rhc_krumholz18} models is consistent with the low dark matter fraction measured in the inner disk, and with the evidence of a centrally driven outflow (Paper~I).

\section{Conclusions} \label{sec:conclusion}

We have obtained deep, kiloparsec-scale resolution ALMA observations in the \cii~158~$\mu$m transition of HZ4, a main-sequence galaxy at $z=5.5$. The results are presented in two papers. In the first one \citep[Paper~I; ][]{rhc_rhc21} we analyzed the interstellar medium, outflow, and halo properties of the system. Here, we focus on the kinematics. The main results can be summarized as follows:

\begin{itemize}
    \item {\bf Evidence for a regular rotating disk at \boldmath{$z=5.5$}:} We analyze the disk kinematics of \name\ based on the 2D kinematic structure and the 1D kinematic profile extracted along the major kinematic axis (Fig.~\ref{fig:mom+RC}). 
    We simultaneously model the 
    rotation curve and the velocity dispersion profile using the fully forward-modeling 3D kinematic code DYSMAL, which takes into account important observational effects such as beam smearing. We measure an intrinsic velocity dispersion of  $\sigma_{0}=65.8^{+2.9}_{-3.3}$~km~s$^{-1}$ and $V_{\rm rot}/\sigma_{0}=2.2$. The latter, combined with the smooth, monotonic velocity gradient observed across the galaxy, the alignment between the morphological and kinematic major axis, and a centrally peaked velocity dispersion profile, strongly suggests that \name\ has a regular rotating disk already in place at $z=5.5$, when the Universe was only $\sim1$~Gyr old.
    
    \medskip
    
    \item {\bf High intrinsic velocity dispersion and the importance of radial transport:} The high intrinsic velocity dispersion ($\sigma_{0}$) measured in \name\ is in agreement with the trends of increasing $\sigma_{0}$ with redshift observed in main-sequence galaxies up to $z\sim4$ (Fig.~\ref{fig:evol}). It is also expected from the high (model-based) cold gas mass fraction measured in the system ($M_{\rm gas}/M_{\star}=(M_{\rm bar}-M_{\star})/M_{\star}=0.7$). To identify the processes responsible for the high level of turbulence in the disk of \name, we use the analytic galaxy evolution models of \cite{rhc_krumholz18} that consider contribution to the disk turbulence by stellar feedback and radial transport. We find that only the models that include radial transport of gas are capable of reproducing the high intrinsic velocity dispersion observed in \name\ (Fig.~\ref{fig:transport}).
    
    \medskip
    
    \item {\bf Low dark matter fraction on galactic scales:} The \cii\ rotation curve in \name\ extends out to 1\arcsec ($\sim6$~kpc), which corresponds to $\sim2\times R_{\rm eff}$ (Fig.~\ref{fig:mom+RC}). The rotation curve peaks at $\sim1.2\times R_{\rm eff}$ and then drops. From the analysis of the contribution of baryons and dark-matter to the intrinsic rotation curve, we find that \name\ is baryon-dominated on galactic scales, with a dark-matter fraction at the effective radius $R_{\rm e}$ of $f_{\rm DM}(R_{\rm e})=0.41^{+0.25}_{-0.22}$ (Fig.~\ref{fig:model}). This low dark-matter fraction is comparable to that found in systems that could be the descendants of \name: massive ($M_{\star}\approx10^{11}~M_{\odot}$), star forming galaxies at $z\approx2$ \cite[e.g,][]{rhc_genzel20}, and passive, early-type galaxies at $z\approx0$ \cite[e.g,][]{rhc_cappellari16}.
    
\end{itemize}

Future spatially-resolved kinematics surveys of the main-sequence galaxy population at $z\sim4-5$ are needed to draw statistically significant conclusions.
This will be possible thanks to observations of the ionized gas with the {\it James Webb Space Telescope}, and upcoming surveys of the cold gas using the \cii\ transition such as the recently approved ALMA Cycle 8 Large Program CRISTAL.\footnote{\url{https://www.alma-cristal.info}} 

\begin{acknowledgements}
We thank the referee for very useful comments
and suggestions that improved the manuscript. R.H.-C. thanks the Max Planck Society for support under the Partner Group project "The Baryon Cycle in Galaxies" between the Max Planck for Extraterrestrial Physics and the Universidad de Concepción. R.H-C also acknowledge financial support from Millenium Nucleus NCN19058 (TITANs) and support by the ANID BASAL projects ACE210002 and FB210003. H{\"U} gratefully acknowledges support by the Isaac Newton Trust and by the Kavli Foundation through a Newton-Kavli Junior Fellowship.
\end{acknowledgements}

%
%

\begin{appendix}

\section{DYSMAL analysis -- MCMC joint posterior distribution} \label{sec:appendix}

\begin{figure*}
\centering
\includegraphics[scale=0.2]{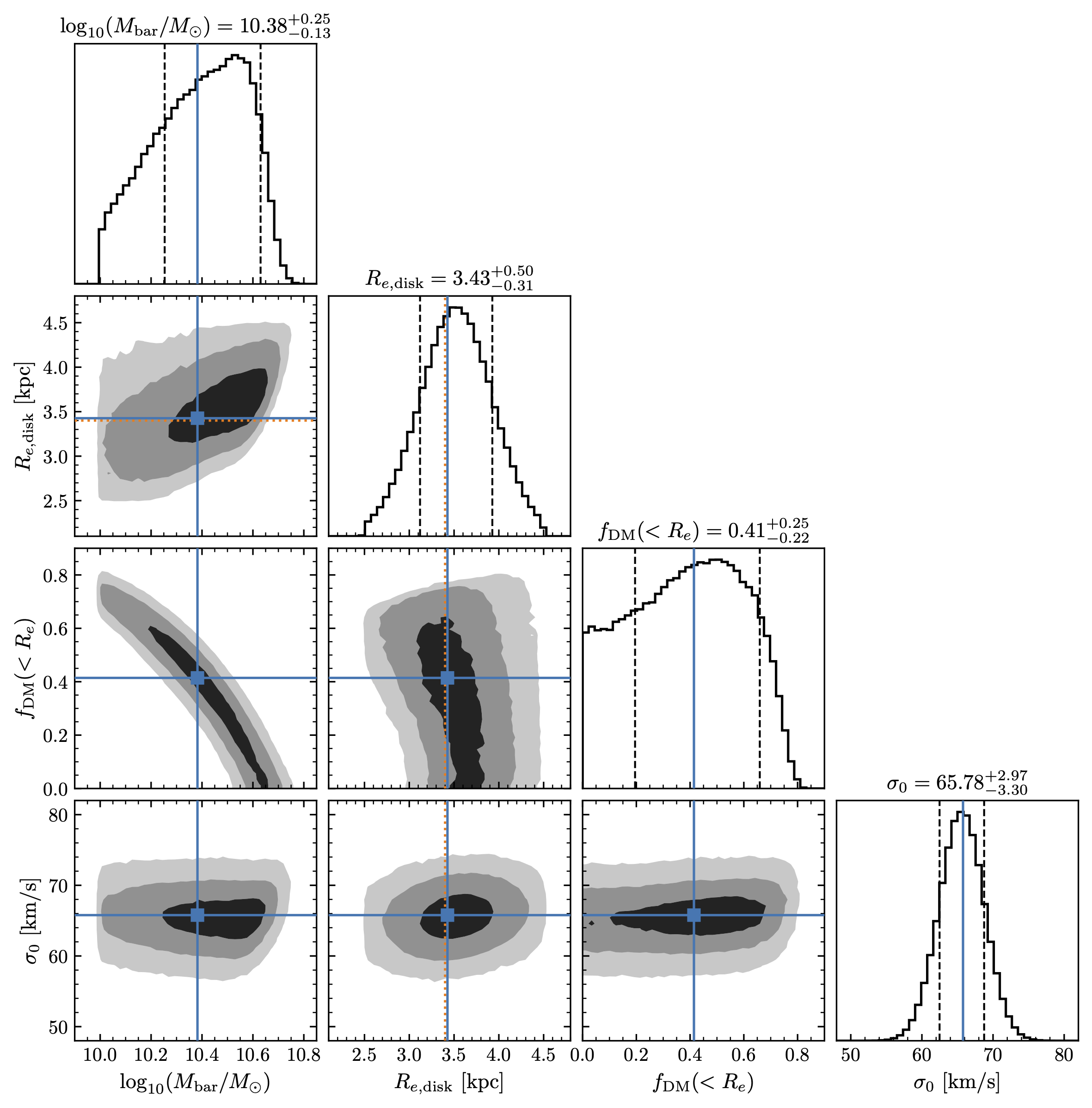}
\caption{MCMC ``corner plot'' for the kinematic modeling of \name. Figure shows the 1D and 2D projections of the posterior probability distributions of the four free parameters: the total baryonic mass ($M_{\rm bar}$), the effective radius ($R_{\rm e}$), the dark matter fraction within one effective radius ($f_{\rm DM}(\leq R_{\rm e})$), and the intrinsic velocity dispersion ($\sigma_{0}$). For Gaussian priors, the centers are marked with orange lines. The maximum a posteriori values of each parameter (found by jointly analyzing the posteriors of all parameters; the ``best-fit’' values) are shown with blue squares and lines. For the 2D histograms, the contours correspond to the 1, 2 and 3$\sigma$ confidence intervals. For the 1D histograms, the uncertainties are the shortest interval enclosing or 1$\sigma$ of the marginalized posterior distribution.
\label{fig:corner}
}
\end{figure*}

As described in Section~\ref{sec:dysmal}, we model the kinematics of \name\ using the DYSMAL code \citep{rhc_cresci09,rhc_davies11,rhc_wuyts16,rhc_price21}. DYSMAL explores the parameter space based on a MCMC sampling using the EMCEE package \citep{rhc_foreman-mackey13}. In our modeling the free parameters are: the total baryonic mass, the effective radius, the dark matter fraction, and the intrinsic velocity dispersion $\sigma_0$. Fig.~\ref{fig:corner} shows the resulting joint posterior probability distributions of the model parameters (or ``corner plot''). 

\section{Position-velocity diagram} \label{sec:pv-appendix}

\begin{figure*}
\centering
\includegraphics[scale=0.25]{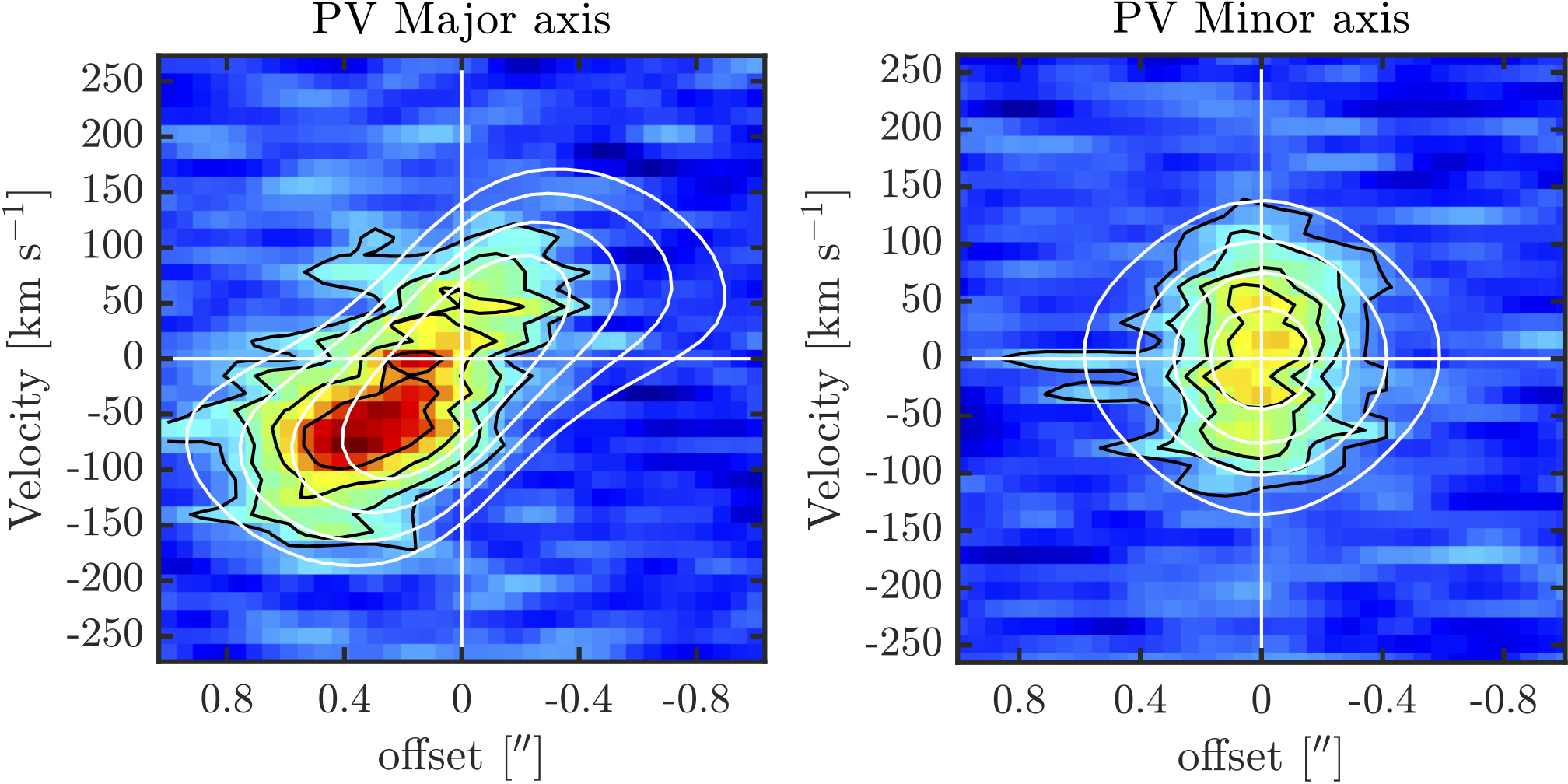}
\caption{{\it(Left)} Position-velocity diagram of \name\ extracted along the major kinematic axis (${\rm PA}=200^{\circ}$) using a pseudo-slit of 0.4\arcsec width. The black contours show the 3, 5, 8 and 10$\sigma$ contours ($1\sigma=0.1$ mJy beam$^{-1}$), and the white contours show the best-fit DYSMAL model at the same contour levels. {\it(Right)} Similar to the left panel, but for the position-velocity diagram extracted along the minor kinematic axis. The black contours show the 2, 4, 6 and 8$\sigma$ contours, and the white contours show the best-fit DYSMAL model at the same contour levels.
\label{fig:pv}
}
\end{figure*}

As discussed in Section~\ref{sec:dysmal}, we use the DYSMAL kinematic code to model the 1D kinematics of \name. Fig.~\ref{fig:pv} shows the position-velocity diagram of \name\ and the resulting best-fit DYSMAL model cube extracted along the major (${\rm PA}=200^{\circ}$) and minor kinematic axes.

\end{appendix} 

\bibliographystyle{aa}
\bibliography{references.bib}

\end{document}